\documentclass[twocolumn,showpacs,floatfix,superscriptaddress,citeautoscript,cite]{revtex4}
\usepackage{graphicx}
\usepackage{xfrac}

\begin{document}

	\author{J. Sivek}
		\email{jozef.sivek@ua.ac.be}
	\author{O. Leenaerts}
		\email{ortwin.leenaerts@ua.ac.be}
	\author{B. Partoens}
		\email{bart.partoens@ua.ac.be}
	\author{F. M. Peeters}
		\email{francois.peeters@ua.ac.be}

	\affiliation{Departement Fysica, Universiteit Antwerpen, Groenenborgerlaan 171, B-2020 Antwerpen, Belgium}

	\title{First-Principles Investigation of Bilayer Fluorographene}

	\date{\today}


\begin{abstract}
\textit{Ab initio} calculations within the density functional theory
formalism are performed to investigate the stability and electronic 	
properties of fluorinated bilayer graphene (bilayer
fluorographene). A comparison is made to previously investigated
graphane, bilayer graphane, and fluorographene.
Bilayer fluorographene is found to be a much more stable
material than bilayer graphane.
Its electronic band structure is similar to that of monolayer fluorographene,
but its electronic band gap is significantly larger (about 1 eV).
We also calculate the effective masses around the $\Gamma$-point
for fluorographene and bilayer fluorographene and find
that they are isotropic, in contrast to earlier reports.
Furthermore, it is found that bilayer fluorographene
is almost as strong as graphene, as its 2D Young's modulus
is approximately 300 $\mathrm{N}\,\mathrm{m}^{-1}$.

\end{abstract}

\maketitle

\section{Introduction}

Since the first reports on the successful isolation of stable
two-dimensional crystals consisting of a single atom layer
by Novoselov {\it et\ al.} in 2004, \cite{novoselov_2004, novoselov_2005}
researchers have been looking for ways to employ these new materials
in real-world applications. Of particular interest is graphene,
a single-layer derivative of graphite. Graphene attracts
a lot of attention with its high crystal quality
and very promising electronic properties. \cite{geim_2007}

Despite its unique characteristics that might
be exploited in future applications,
\cite{tombros_2007, novoselov_2005-dirac_fermions, bao_2011, lui_2011}
there exists a serious obstacle to use graphene in electronics as we
know it today. Graphene is a zero-gap semiconductor, and the absence of
a band gap is protected by inversion and time-reversal
symmetry. \cite{kane_2005} Several approaches have been
used or proposed to open and control
a band gap in graphene and graphene derivatives:
(i) mechanical modification, such as cutting
the graphene sample into nanoribbons, \cite{han_2007, li_2008} 
(ii) application of a potential difference over a bilayer of 
graphene, \cite{ohta_2006, mccann_2006, samarakoon_2010}
and finally, (iii) chemical modification that confines the $p_z$ 
electrons into covalent bonds.
\cite{nair_2010, elias_2009, cheng_2010, withers_2010, zboril_2010-f_graphene_and_derivatives}
The last approach includes the adsorption of radicals,
such as oxygen, hydrogen, and fluorine atoms, on the surface of graphene
that form chemical bonds with the carbon atoms and change
their hybridization from $\mathrm{sp}^{2}$ to $\mathrm{sp}^{3}$.
The chemical derivatives of graphene preserve the
two-dimensional character of graphene, but
they have vastly different electronic properties. This makes them 
possibly more interesting materials for applications in 
electronics than intrinsic graphene.

Among the possible derivatives of graphene, there are two that have
attracted special attention, namely, graphane and fluorographene
(graphene fluoride), which result from the hydrogenation and
fluorination of graphene, respectively. They are theoretically
predicted \cite{sluiter_2003,sofo_2007,artyukhov_2010, 
ortwin_2010-fh_graphene, samarakoon_2011-fluorographene,
sahin_2011, garcia_2011-graphane-like_sheets, ueta_2012}
and experimentally observed
\cite{nair_2010, cheng_2010, jeon_2011, zboril_2010-f_graphene_and_derivatives, elias_2009, luo_2009}
to form crystalline materials, in contrast to, for example, graphene oxide.
\cite{dikin_2007, eda_2010}

These new two-dimensional crystals are currently the subject 
of a considerable amount of experimental and 
theoretical work.
However, the process of chemical modification is not restricted to
monolayer graphene. It has been proposed that this process can 
be expanded to bilayer graphene as well. \cite{ortwin_2009-h_bi_graphene}
\textit{Ab initio} calculations showed that the weak van
der Waals forces between the graphene layers are replaced by much  
stronger covalent bonds that stabilize the structure and that, at full coverage, 
a bilayer analogue of graphane is formed. The electronic 
structure of intrinsic monolayer and bilayer graphane are very similar,
\cite{ohta_2007} but their mechanical/elastic properties are
expected to be substantially different.

The difference in elastomechanical qualities can be an important issue.
It was recently demonstrated \cite{neek_2011} that the roughness 
of monolayer graphane surfaces is considerably larger than
that of graphene. This increase of the size of the ripple formation  
can be explained by the lower stiffness of graphane \cite{munoz_2010}
and its different vibrational properties as compared with
graphene. \cite{hartwin_2011}
One way to reduce the increased roughness is to consider
bilayers instead of single layers. As is well known from experiment,
the ripple formation in bilayer graphene is strongly reduced;
that is, they are removed by the interlayer interaction. \cite{meyer_2007}
A similar effect can be expected for bilayer graphane where
the interlayer interaction is even more important. 

\begin{figure}[h]
  \centering
\includegraphics[width= 3.25 in]{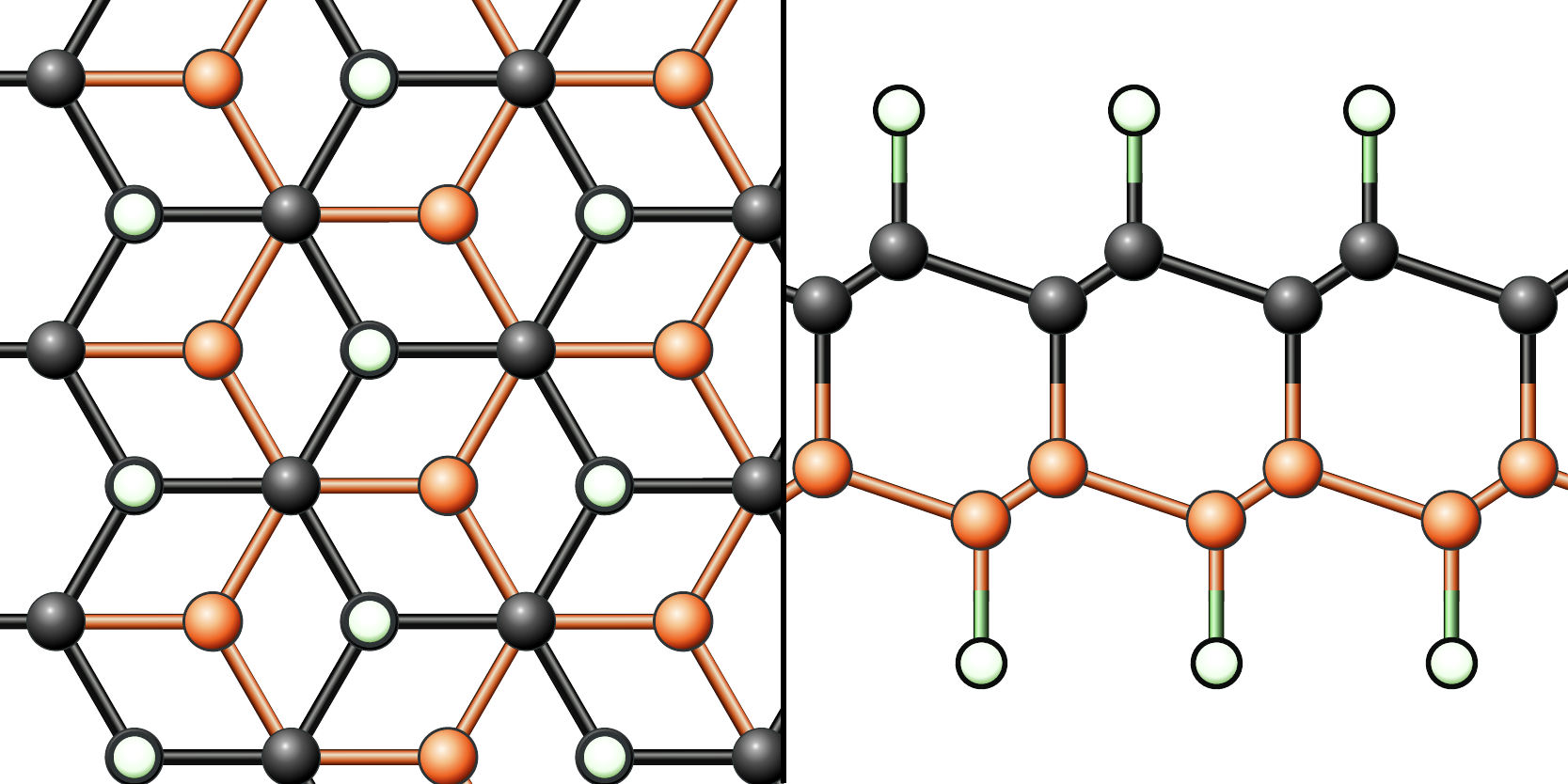}
	\caption{\label{fig_fluorographene}
	Top (left) and side (right) view
	of bilayer fluorographene. The carbon atoms of
	the two layers are given by different colors (shades)
	for clarity and the fluorine atoms are in light green
	(open circles with the smallest diameter).
	}
\end{figure}

In this paper, we investigate whether it is also possible
to fluorinate bilayer graphene.
We perform {\it ab initio} calculations to determine the
stability and structural properties of a fluorinated
bilayer of graphene, hereafter called bilayer fluorographene
(for a better notion of the chemical composition
and the structure, see Fig.\ \ref{fig_fluorographene}).
We find that bilayer fluorographene is a much more
stable compound than bilayer graphane, although
more fluorination is needed to induce interlayer covalent
C--C bonds. The structural properties of bilayer fluorographene
are in between those of monolayer fluorographene and diamond;
other characteristics are closer to those of initial fluorographene.

This paper is organized as follows: 
First, we describe the computational details of our {\it ab initio}
calculations, followed by an investigation of the stability and
the formation conditions of interlayer covalent C--C bonds. Next, the overall
stability and the geometrical properties of fully fluorinated bilayer
graphene (i.e.,\ bilayer fluorographene) are examined.
These properties are subsequently compared with those of
diamond and fluorographene. To conclude, we investigate
the electronic band structure and band gaps of single-layer
and bilayer fluorographene, together with
the effective masses of the various possible charge carriers
and elastic properties represented by their 2D Young's moduli.

\section{Calculations}

All our calculations were done within the density functional theory (DFT)
formalism as implemented in the VASP package with usage of the local density
approximation (LDA) and the Perdew, Burke, and Ernzerhof \cite{perdew_1996-gga-pbe}
generalized gradient approximation (GGA) for the exchange-correlation functional.
We made use of the projector augmented wave method \cite{bloch_1994-paw}
and a plane-wave basis set with an energy cutoff of 500 eV.
The relaxation of atomic positions
was performed with forces smaller than 0.01~eV~\AA$^{-1}$.

Three types of supercells were used in our calculations: 
$3\times3$ and $2\times2$ supercells to study the adsorption properties
of fluorine on a graphene bilayer for different concentrations and configurations
of fluorine and a $1\times1$ unit cell for the calculation
of the properties and electronic band structure of fully
fluorinated graphene and bilayer graphene. 

The sampling of the Brillouin zone was done for the different
supercells with the equivalent of a $24\times24\times1$ Monkhorst--Pack \cite{monkhorst_1976}
$k$-point grid for the monolayer or bilayer graphene unit cell
(containing two carbon atoms per layer). Spin polarization was not
included in the calculations because fluorination is not expected to
induce magnetism in graphene. \cite{sofo_2011}

Because periodic boundary conditions were applied in all three dimensions,
the height of the supercell was set to 20~{\AA} to include enough vacuum 
to minimize the interaction between adjacent layers.
Additionally, we have performed a convergence test with respect to
the planar size of the supercell. We obtain an accuracy
in binding energies of less than 0.1~eV and in bond lengths
of less than 0.01~{\AA} with the chosen $3\times3$ supercell.

\section{Results and Discussion}

\subsection{Fluorination of Bilayer Graphene}

The main objective of this paper is to investigate
chemisorption of fluorine atoms on both sides
of bilayer graphene. Hydrogenation and fluorination
of the carbon atoms of graphene change
their hybridization from $\mathrm{sp}^{2}$ to $\mathrm{sp}^{3}$.
In the case of hydrogenation of bilayer graphene,
it was demonstrated that stable interlayer C--C bonds
are induced at high concentrations
of adsorbates.\cite{ortwin_2009-h_bi_graphene} In this section,
we perform a similar study for the case of fluorination
and highlight the similarities and differences as compared
to hydrogenated bilayer graphene. We make use of
the local density approximation, because this approximation
gives a better description of the interlayer interaction
in bilayer graphene over the generalized gradient approximation.\cite{hasegawa_2004}

\begin{figure}[h]
  \centering
\includegraphics[width= 2.20 in]{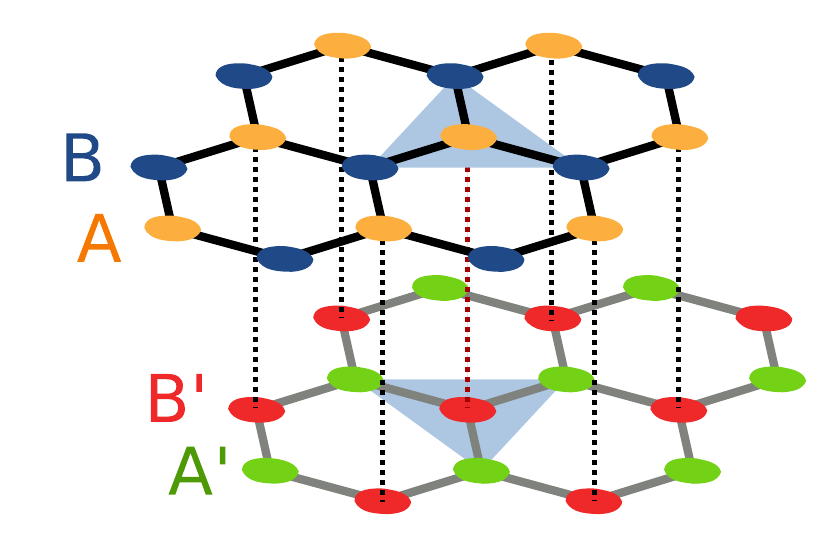}
	\caption{\label{fig_sublattices}
	Bilayer of AB-stacked graphene with the four different
	sublattices indicated by different colors.
	The A and B' sublattices are situated on top of each other, as illustrated by
	the dotted lines. The carbon atoms at the corners of the light
	triangles indicate the six neighbors to be fluorinated.
	}
\end{figure}

We only consider chemisorption on bilayer graphene
with AB stacking because this is the most energetically
favorable way of stacking graphene layers. 
As indicated in Fig.\ \ref{fig_sublattices}, this  gives
rise to four sublattices (two in each layer) of which two,
namely, the A and B' sublattice, are exactly aligned
on top of each other while the sublattices B and A'
are not. Since the adsorbates under investigation always
chemisorb on top of a single C atom, there are two 
(inequivalent) adsorption sites, namely, on the A (or B')
sublattice and on the B (or A') sublattice. 
It was found in an earlier study of hydrogen adsorption on
bilayer graphene that there is a difference in adsorption
energy of approximately 0.03 eV in favor of the B sublattice.
\cite{ortwin_2009-h_bi_graphene} The H atoms are, therefore, likely to 
attach to those carbon atoms that have no direct neighbors in 
the other layer.
For F atoms, we find the same preference for adsorption on the B
sublattice. However, the difference in adsorption energy for
the two sublattices is only 0.3 meV, which is about 100 times
smaller than in the case of hydrogen. 
This is caused by the different nature of C--F 
bonds as compared with C--H bonds: C--H bonds are purely covalent, 
but C--F bonds have a partially ionic character because of the 
large difference in electronegativity between C and F.

The formation of a chemical bond between an adsorbate and a C 
atom of the bilayer of graphene induces a change in the
hybridization of this C atom from $\mathrm{sp}^{2}$ to $\mathrm{sp}^{3}$. 
If the concentration of attached adsorbates on the preferred
sublattice increases, the carbon atoms of the other sublattice 
also change their hybridization to $\mathrm{sp}^{3}$. This allows for 
the formation of covalent bonds between the two graphene layers.
These interlayer bonds are formed between
carbon atoms of the A and B' sublattices that are pushed to
one another because of the changed hybridization of their neighbors.

For a single interlayer C--C bond, there are six neighbors: three
atoms in the B and three in the A' sublattice (Fig.\ \ref{fig_sublattices}).  
However, it is not necessary for all of these neighbors to change
their hybridization in order for the interlayer bond to be formed,
as in the case of hydrogen chemisorption on bilayer graphene,
where only four of the neighboring carbon atoms need to be hydrogenated
to induce a covalent interlayer bond. \cite{ortwin_2009-h_bi_graphene}

To study the formation of interlayer bonds upon
fluorination in more detail, we gradually increase
the number of attached fluorine atoms on neighboring
carbon atoms from one to six. We completely relax
the resulting systems and calculate the average and minimum
distances between the C atoms of the two layers. The results of these
computations are summarized in Table \ref{table_3x3}.
From this table, a gradual decrease of the interlayer
distance can be seen upon increasing the concentration
of F atoms. At higher concentrations, however, a substantial
jump in the minimum distance, $d_{\mathrm{min}}$, occurs when going
from four to five fluorinated neighbors. This jump is about
1~{\AA} in size and clearly indicates the formation
of an interlayer bond. It is thus necessary to fluorinate
five of the six neighbors to induce an interlayer bond, in contrast to
hydrogenation, where four hydrogenated neighbors suffice.

\begin{table}[h]

	\caption{
	Interlayer chemical bond formation in a $3\times3$ supercell.
	The binding energy per F atom ($E_{\mathrm{b}}$), the formation energy per F atom
	($E_{\mathrm{f}}$) and the average ($d_{\mathrm{avg}}$) and
	minimum distance ($d_{\mathrm{min}}$) between the C atoms of the two layers are
	shown for different number of adsorbed fluorine atoms and different
	configurations. Presence of interlayer bond is marked as chemical bond.
	Energies are given in eV and distances~in~\AA.
	\label{table_3x3}}

\begin{tabular}{ c l c c c c c }
	\hline \hline
	no.          & configuration & $E_{\mathrm{b}}$  & $E_{\mathrm{f}}$ &$d_{\mathrm{avg}}$&$d_{\mathrm{min}}$& chemical \\
	F atoms      &               &        &          &         &         & bond  \\
	\hline
	1            & B             & -2.525 & -0.819 & 3.326   & 3.256   & no  \\
	2            & BB            & -2.405 & -0.700 & 3.326   & 3.190   & no  \\
	2            & BA'           & -2.555 & -0.850 & 3.213   & 3.085   & no  \\ 
	3            & BBB           & -2.290 & -0.585 & 3.332   & 3.126   & no  \\
	3            & BBA'          & -2.487 & -0.782 & 3.176   & 2.979   & no  \\
	4            & BBBA'         & -2.384 & -0.679 & 3.150   & 2.903   & no  \\
	4            & BBA'A'        & -2.479 & -0.773 & 3.061   & 2.821   & no  \\
	5            & BBBA'A'       & -2.450 & -0.745 & 2.678   & 1.757   & yes \\
	6            & BBBA'A'A'     & -2.560 & -0.854 & 2.598   & 1.737   & yes \\
	\hline \hline
\end{tabular}
\end{table}

Additional information about the stability of the interlayer bond
can be obtained by examining the formation and binding energy of
the system at different concentrations of adsorbed F atoms. We define
the formation energy ($E_{\mathrm{f}}$) as the energy per attached fluorine atom
with respect to intrinsic bilayer graphene and the diatomic molecule
F$_2$. The binding energy ($E_{\mathrm{b}}$), on the other hand, is defined
as the energy per fluorine atom (or CF pair) with respect
to intrinsic bilayer graphene and atomic fluorine.
Both energies are given in Table \ref{table_3x3}. 

The formation energy, $E_{\mathrm{f}}$, is negative in all cases,
which means that all the configurations are stable
against molecular desorption from the graphene surface.
This should be contrasted to the same concentrations of 
hydrogen atoms on the surface of bilayer graphene,
where the energy becomes negative only
in the case of almost fully hydrogenated bilayer
graphene.\cite{ortwin_2009-h_bi_graphene}
The value of the formation energies in the case of
fluorination is significantly lower (almost 1 eV)
than in the case of hydrogen chemisorption.
This can be attributed to the weaker bond in the
F$_2$ as compared with the H$_2$ molecule and is similar
to the previously studied case of monolayer graphane
and fluorographene.\cite{samarakoon_2011-fluorographene}
The calculated binding energies also show that it is
energetically favorable for the fluorine atoms to attach
themselves on both sides of the bilayer: for the same number
of chemisorbed F atoms, the configuration in which
these atoms are distributed as equally as possible
between the two layers is lower in energy and thus more stable.
The fact that adsorbed F atoms on one side of the bilayer
make it favorable for other F atoms to adsorb
on the other side increases the chance of interlayer bond
formation. Because these bonds are stable, we can imagine
this process of fluorination and interlayer bond
formation to continue until a fully covered bilayer
of graphene fluoride is formed.

\begin{figure}[h]
  \centering
\includegraphics[width= 3.25 in]{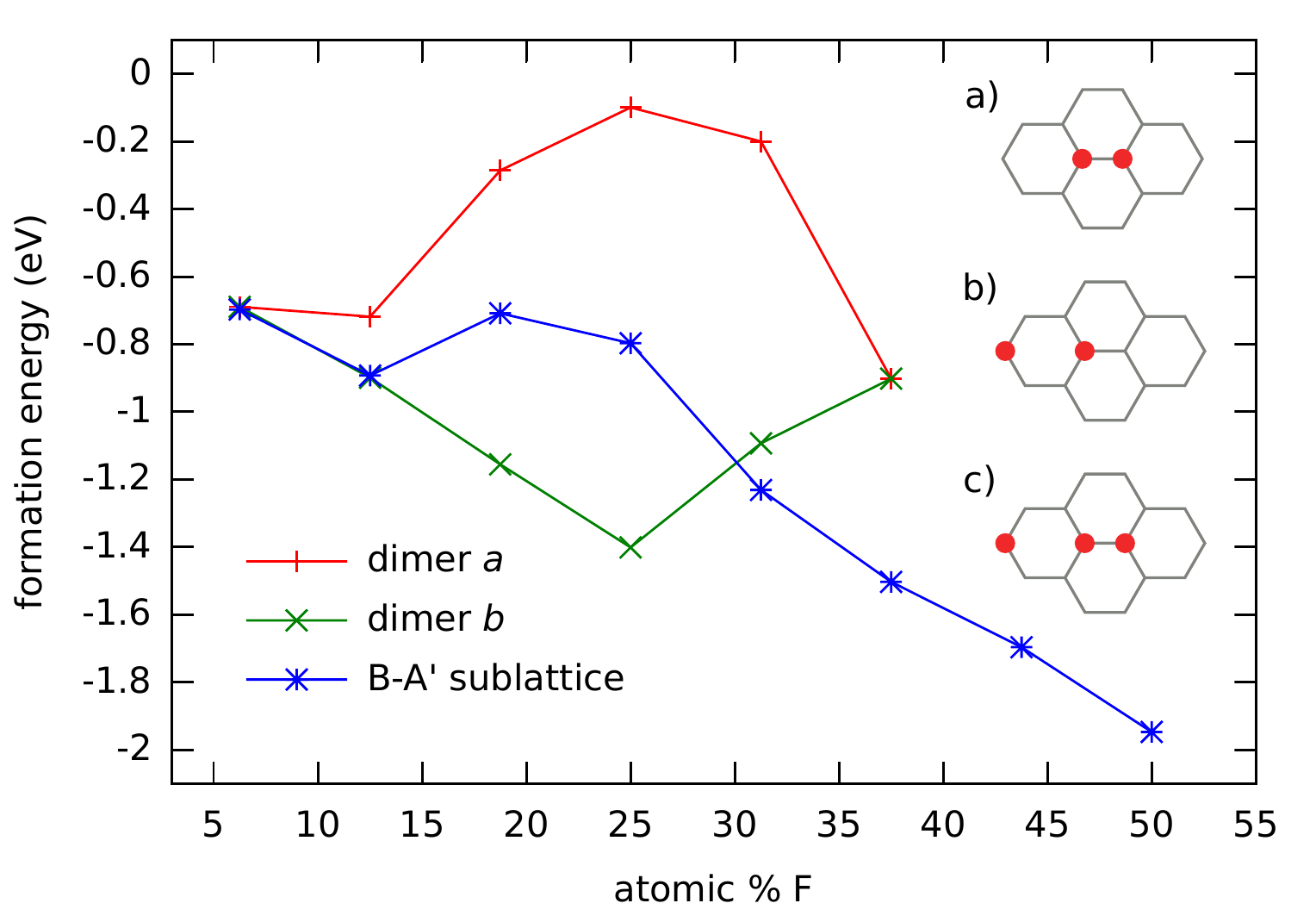}
	\caption{\label{fig_formation_energies}
	Gradual fluorination of a $2\times2$ supercell of bilayer graphene.
	For increasing F concentration the formation energy per
	fluorine atom is shown. The curves show
	different distributions of the F atoms: according to dimer
	configuration \emph{a} (inset a) and \emph{b} (inset b), only on
	sublattices B and A', and the placement of the F atoms only on
	sublattices A and B'. Dimers \emph{a} and \emph{b} can not be preserved for 
	higher levels of fluorination; instead they transform
	into the trimer configuration (inset c).
	}
\end{figure}

To test if the aforementioned interlayer bond creation
scenario is truly energetically favorable, we considered
a $2\times2$ supercell for a gradual fluorination with
different configuration patterns. To be able to choose
from the large number of different possible configurations,
we need some guidelines for further investigation.
As a first guideline, we distribute the F atoms equally 
on both sides of the bilayer because it is
more energetically favorable, as discussed above.
Four prevalent configurations are known for the fluorination of graphene:
chair, boat, zigzag, and armchair. The chair configuration
can be readily extended to the fluorination of bilayer 
graphene discussed above. The remaining three configurations share as a common feature
that they contain dimers, as depicted in the inset of Fig. \ref{fig_formation_energies}.
Those dimers are used as the second guideline
to reduce configuration space.
The dimer \emph{a} is composed of F atoms bonded with
directly neighboring C atoms, whereas the dimer \emph{b} consists
of F atoms bonded with distant C atoms on the opposite sides of 
the hexagonal ring. For higher fluorine concentrations
(above two dimers, one at each side of the bilayer), dimers \emph{a} and \emph{b} are combined
and create a trimer configuration, as depicted in the inset c of Fig. \ref{fig_formation_energies}.

The calculated results are given in Fig. \ref{fig_formation_energies}.
In this figure, we present the formation energy per adatom for the different
fluorination pathways. These results show that,
for low concentrations of F atoms, the dimer \emph{b}
configuration is more stable (carbon atoms from both sublattices
are fluorinated, and no interlayer bonds are created).
However, higher concentrations of F atoms
are not feasible with the dimer configurations
as the formation energy increases. However, if only the B and A' sublattices
are fluorinated, the formation energy
can decrease even more, finally creating bilayer fluorographene, with completely
saturated C atoms, as can be seen in Fig.\ \ref{fig_fluorographene}.
It is also important to point out that the total formation energy
for all adsorbed F atoms (not shown in the figure)
is a monotonically decreasing function
of the F atom concentration, and therefore, the local minimum
for dimer \emph{b} will not result in an interruption of the adsorption.

Our calculations also showed that, for both dimer configurations,
no interlayer bonding was formed. The dimers \emph{a} and \emph{b}
transform for higher coverage into the trimer configuration, as indicated by
the connection of the plots. For the adsorption on B and A' sublattices,
the interlayer bond is created already with only four F atoms.
This apparently different result from the previous $3\times3$ supercell
calculations is due to the choice of the size of supercell.
Four adatoms in a $2\times2$ supercell correspond to a higher
percentage of F per C atoms than five adatoms
in the $3\times3$ supercell. 

We also performed calculations
for the case when the A and B' sublattices, lying on top
of each other, were fluorinated. As already
mentioned, the preference for B (A') over
A (B') sublattice for a single adatom is very small and
unlikely to conduct the adsorption pattern.
With increased F atom coverage, the
formation energy per adatom for fluorination of the A and B' sublattice
was found to be an increasing  function of the total coverage.
This results in a very unfavorable alternative
over fluorination of B and A' sublattices.

From the calculations on the F adsorption in
a $2\times2$ supercell, we can conclude that
the chair configuration is the lowest-energy configuration,
the only configuration that leads to interlayer bonding and
in which a high fluorine concentration can be reached.

\subsection{Properties of Bilayer Fluorographene}

In this section, we examine the properties of bilayer
fluorographene in more detail and compare
them with those of monolayer fluorographene and diamond. 
For the study of the geometrical properties,
we include also GGA calculations because they give
more accurate bond lengths and angles. 
A summary of the geometrical and electronic properties is given in
Tables \ref{table_measurements} and \ref{table_energies}.
The results for monolayer fluorographene compare well with
previous calculations on this system.\ \cite{artyukhov_2010}
As can be expected, the cell size and bond lengths and
angles of bilayer fluorographene have values between
those of monolayer fluorographene and diamond.
Overall, the geometrical properties are close to the one
of bulk diamond due to the same hybridization
of the carbon atoms.\ \cite{garcia_2011-graphane-like_sheets}

The calculated C--C bond lengths are larger
than the ones in diamond. This difference can be explained
from a chemical point of view as a result
of the depopulation of the bonding orbitals between the carbon atoms after
substantial charge transfer to the F atom,
which is the result of the difference
in electronegativity between C and F atoms.

Additionally, we have observed the shortening of the C--F bond
with increasing amount of fluorination on both sides of bilayer
fluorographene. In a $3\times 3$ supercell, the C--F bond length varies
from 1.50 {\AA} for configuration B to 1.43 {\AA} for
configuration BBBA'A'A'. As can be seen from Table \ref{table_measurements},
the C--F bond length decreases even more to 1.38 {\AA} for fully
fluorinated bilayer graphene. The source of this behavior
can be attributed to the ionic character of the C--F bond and a gradual
transformation from $\mathrm{sp}^{2}$ to $\mathrm{sp}^{3}$ hybridization of the C atoms.\ \cite{sofo_2011}

\begin{table}[h]

	\caption{
	Properties of single layer and bilayer fluorographene: the unit cell
	length ($a$), the distances ($d$) and angles ($\theta$) between
	neighboring atoms. Distances are given in {\AA} and angles in deg.
	\label{table_measurements}}
	
\begin{tabular}{l cc | cc | cc}
	\hline\hline
	   & \multicolumn{2}{c|}{fluorographene}  & \multicolumn{2}{c|}{bilayer}        & \multicolumn{2}{c}{diamond} \\
	   & \multicolumn{2}{c|}{}                & \multicolumn{2}{c|}{fluorographene} & \multicolumn{2}{c}{} \\
	
	                          &  LDA  &  GGA  &  LDA  &  GGA  &  LDA  &  GGA  \\
	\hline
	$a$                       & 2.555 & 2.596 & 2.525 & 2.563 & 2.499 & 2.527 \\
	$d_{\mathrm{C-C}}$        & 1.553 & 1.576 & 1.541 & 1.563 & 1.531 & 1.547 \\
	$d_{\mathrm{C-C'}}$       & n/a   & n/a   & 1.537 & 1.554 & 1.531 & 1.547 \\
	$d_{\mathrm{C-F}}$        & 1.365 & 1.382 & 1.361 & 1.377 & n/a   & n/a   \\
	$\theta_{\mathrm{CCC}}$   & 110.7 & 110.9 & 110.0 & 110.2 & 109.5 & 109.5 \\
	$\theta_{\mathrm{CCC'}}$  & n/a   & n/a   & 108.9 & 108.7 & 109.5 & 109.5 \\
	$\theta_{\mathrm{CCF}}$   & 108.2 & 108.0 & 108.9 & 108.7 & n/a   & n/a   \\
	
	\hline\hline
\end{tabular}
\end{table}

The electronic properties of monolayer and bilayer
fluorographene are given in Table\ \ref{table_energies}: we calculated
the band gap of these materials together with the formation energy
per atom (in contrast to previously used formation energies per fluorine atom).
For comparison, the values of these quantities are also given for diamond.

The calculated band gap for monolayer fluorographene is in good agreement
with previously published results when using GGA calculations. \cite{lu_2009}
Although the computed band-gap value for fluorographene is close
to experimentally measured values\ \cite{nair_2010}, this should be seen
as a coincidence. LDA and GGA calculations are
known to largely underestimate the value of the band gap.
The calculated band gaps are, in fact 2 times lower
than more accurate results of many-electron GW calculations.\ \cite{samarakoon_2011-fluorographene}
This apparent discrepancy is attributed to the presence
of a considerable amount of defects in the experimental
samples that induce midgap states (similar to defected
graphane\ \cite{berashevich_2010}).

The band gap of bilayer fluorographene is found to be larger than the
one of monolayer fluorographene by approximately 1 eV. This is different from
the case of hydrogenated graphene, where monolayer graphane is found to have a 
slightly larger band gap than bilayer graphane. \cite{ortwin_2009-h_bi_graphene}
In fact, within LDA and GGA calculations, graphane has a larger band gap
than monolayer fluorographene,\cite{tang_2011, tang_2010-bn} but this observation does not apply
to bilayer compounds where bilayer fluorographene surpasses
bilayer graphane in the size of the energy gap.
Although our calculations are not accurate enough to provide the
real band gap, it is probable that this difference in the
size of the band gap between monolayer and bilayer fluorographene
is qualitatively correct. This follows from the fact that
LDA and GGA usually produce correct trends in the variation of the 
band gap among similar systems. Therefore, the band gap of bilayer fluorographene
has been found to be intermediate between that of fluorographene and
that of bulk diamond. In this sense, diamond can be considered as the limit
of multilayer fluorographene. This tendency was observed before:
for theoretically proposed graphite fluoride materials involving
carbon atoms with $\mathrm{sp}^{3}$ hybridization.\ \cite{takagi_2002}

The absolute value of the formation energy of bilayer
fluorographene is smaller than the one of the monolayer, but
still larger than the one of diamond. The main reason for the
observed weakening of the formation energy is the drop of the ratio
of the amount of fluorine atoms per carbon atom in going from monolayer 
to bilayer fluorographene. Overall, the qualitative image
of the stability of the fluorinated materials
corresponds to that of graphane and bilayer graphane in terms of
the aforementioned formation energy and the presence of covalent bonds between
the graphene layers, which stabilize the structure.
Nevertheless, there is a large quantitative difference;
the fluorinated materials are much more stable structures
than the hydrogenated ones.\ \cite{sofo_2007}

\begin{table}[h]
  
	\caption{
	Electronic band gap $E_{\mathrm{gap}}$ and formation energy per atom ($E_{\mathrm{f}}/atom$)
	of monolayer and bilayer fluorographene and diamond
	using LDA and GGA calculations. All the energies are given in eV.
	\label{table_energies}}
	
\begin{tabular}{l cc | cc | cc}
	\hline\hline
	   & \multicolumn{2}{c|}{fluorographene}  & \multicolumn{2}{c|}{bilayer}        & \multicolumn{2}{c}{diamond} \\
	   & \multicolumn{2}{c|}{}                & \multicolumn{2}{c|}{fluorographene} & \multicolumn{2}{c}{} \\
	
	                         &  LDA  &  GGA  &  LDA  &  GGA  &  LDA  &  GGA  \\
	\hline
	$E_{\mathrm{gap}}$       & 2.960 & 3.089 & 3.951 & 4.040 & 5.618 & 5.572 \\
	$E_{\mathrm{f}}/atom$    &-1.057 &-0.944 &-0.722 &-0.593 &-0.011 & 0.132 \\
	
	\hline\hline
\end{tabular}
\end{table}

\begin{figure}[h]
  \centering
\includegraphics[width= 3.25 in]{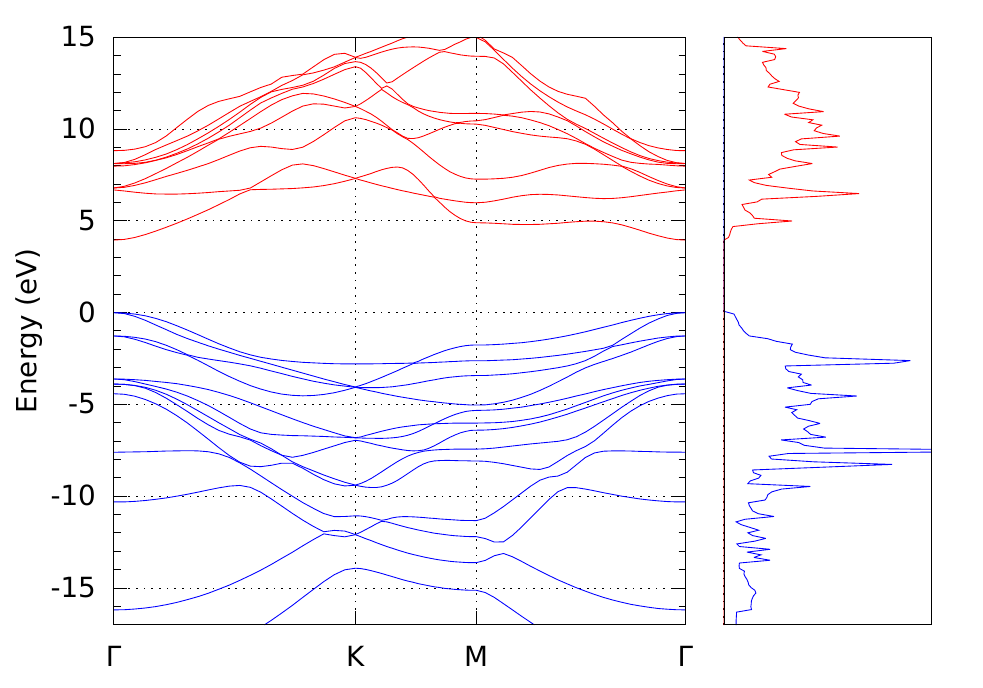}
	\caption{\label{fig_bands}
	Electronic band structure and density of states
	of bilayer fluorographene.
	The energies are relative to the Fermi level ($E_F$=0).
	}
\end{figure}

The band structure of bilayer fluorographene
is displayed in Fig.\ \ref{fig_bands} together with
a diagram of the density of states.
The depicted band structure is seen to be qualitatively
similar to the one of monolayer fluorographene \cite{ortwin_2010-fh_graphene}
with the size of the band gap as the only obvious difference.
We also calculated the effective masses of the charge carriers
around the $\Gamma$-point in order to find other
differences or similarities between the two materials.

The obtained effective masses of electrons, and heavy and light holes
for both studied materials, can be found in Table\ \ref{table_effective_masses}.
It should be noted that these masses were calculated
within the DFT formalism with usage of the LDA for
the exchange-correlation functional, which has been shown
to give reasonable results.\ \cite{oshikiri_2002-effective_mass}

\begin{figure}[h]
  \centering
\includegraphics[width= 3 in]{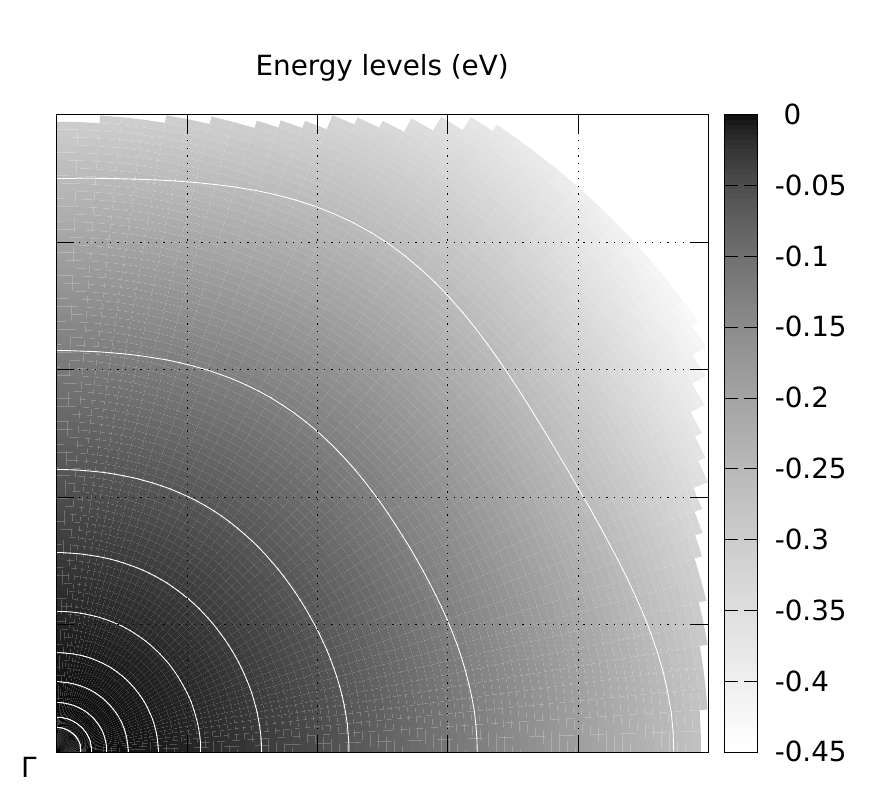}
	\caption{\label{fig_last_valence_band}
	Energy levels of the highest valence band of monolayer
	fluorographene in reciprocal space. The displayed region is a
	squared partition of the Brillouin zone with a side length 0.22~of
	the $\Gamma$--$M$ distance. The $\Gamma$-point is taken as the origin,
	and the $x$ axis directed toward the $M$-point. Contour lines
	are displayed for a better notion of the symmetry.
	}
\end{figure}

The effective mass of charge carriers in crystalline materials
usually depends  on the direction in reciprocal space.
Therefore, we have chosen two common directions for the hexagonal lattice
at the $\Gamma$-point, namely, $\Gamma \rightarrow K$ and $\Gamma \rightarrow M$.
The results for these two directions were found to be
indistinguishable, and so we can conclude that the effective masses
at the $\Gamma$-point are isotropic.
The direction independence for the effective
masses that we observe contradicts previous results in which
the effective masses of monolayer fluorographene were 
found to be highly anisotropic.\ \cite{liang_2011}
Our statement about the isotropic character of the
effective masses is supported by a direct
plot of the energy value map around the $\Gamma$-point.
In Fig.\ \ref{fig_last_valence_band} we show a picture of the highest
valence band of monolayer fluorographene. The isotropic character 
of this band is clearly visible close to the $\Gamma$-point.
Farther away from the $\Gamma$-point an anisotropy
of the surface map is visible, which can be attributed to
fourth- and higher-order effects. However, these higher-order effects
do not induce anisotropy in the effective mass,
which is a second-order effect. Similar results
are found for the light hole and the electrons.

When we compare the effective masses of monolayer and bilayer
fluorographene, we observe only a small difference in its values.
We, therefore, conclude that both materials have similar electronic 
properties apart from a difference in the band gap of about 1 eV.

\begin{table}[h]
  
	\caption{
	Effective masses of holes and electrons (in units of the electron mass)
	and the 2D Young's modulus, E, along the cartesian axes. E is expressed in $\mathrm{N}\,\mathrm{m}^{-1}$.
	\label{table_effective_masses}}
	
\begin{tabular}{ l c c }
	\hline\hline
	              & fluorographene & bilayer        \\
				  &                & fluorographene \\	
	\hline
	$m$           & 0.48           & 0.50  \\
	$m_{hh}$      & 1.13           & 1.10  \\
	$m_{lh}$      & 0.41           & 0.37  \\
	
	$E_{x}$       & 195            & 284  \\
	$E_{y}$       & 197            & 293  \\
	
	\hline\hline
\end{tabular}
\end{table}

In addition to the electronic properties, we calculated the 
(2D) Young's modulus, $E$, for monolayer and bilayer fluorographene, which are also included
in Table\ \ref{table_effective_masses}. We followed the
same approach as applied before in Ref.~~\citenum{ortwin_2010-fh_graphene}.
The 2D Young's modulus of graphene is found to be 307 $\mathrm{N}\,\mathrm{m}^{-1}$
shifted from the experimental value, $E_{exp} = 340 \pm 50\, \mathrm{N}\,\mathrm{m}^{-1}$,\ \cite{lee_2008-elestic_prop_graphene}
and other theoretical values.\ \cite{munoz_2010}
The 2D Young's modulus of fluorographene is found to be \sfrac{1}{3} smaller in comparison
to that of intrinsic graphene, whereas the obtained moduli for bilayer
fluorographene reach almost the values of graphene, making these materials
very strong. The values unveil the isotropic character of the Young's modulus
for both compounds.

\section{Conclusions}

As an extension to previous work,\cite{ortwin_2009-h_bi_graphene, ortwin_2010-fh_graphene}
we studied the potentially interesting case of bilayer fluorographene.
We demonstrated that fluorination of bilayer graphene
results in more stable structures than hydrogenation. This
can be clearly observed by a comparison of the formation energies
of the final structures\ \cite{sofo_2007}
and is accentuated by the fact that the formation energy
of partially fluorinated bilayer graphane is always negative,
in distinct contrast to partially hydrogenated bilayer
graphene. The creation of interlayer chemical bonds occurs at higher
amounts of fluorination as compared with hydrogenation.

The calculated band gap of bilayer fluorographene shows a 30\% increase
over the one of bilayer graphane, and we also observed quantitatively significant
differences between monolayer and bilayer fluorographene.
From the value of the Young's modulus, we can conclude that
bilayer fluorographene is substantially stronger than
monolayer fluorographene and is almost as strong as graphene.

\begin{acknowledgments}
This work is supported by the ESF-Eurocores program EuroGRAPHENE (project CONERAN) and the Flemish Science Foundation (FWO-Vl).
\end{acknowledgments}

\end{document}